\documentclass[journal]{IEEEtran}

\usepackage[adjust,compress]{cite}

\usepackage[normalem]{ulem}

\usepackage{graphicx,amsmath,amssymb}
\usepackage{url}
\usepackage{subfigure}
\usepackage{color}
\usepackage{bm}

\DeclareMathOperator{\rank}{rank}

\newcommand{\set}[1]{\mathrm{#1}}

\newcommand{\ie}{{\it i.e.\/}, }

\providecommand*{\mrm}[1]{\mathrm{#1}}

\newcommand{\lexp}[1]{\mathrm{e}^{#1}}

\newcommand{\We}{W_{\mrm{e}}}

\newcommand{\Wm}{W_{\mrm{m}}}

\newcommand{\ju}{\mathrm{j}}
\newcommand{\diff}{\mathop{\mathrm{\mathstrut{d}}}\!}%Diff op.

\newcommand{\rP}{P_\mrm{rad}}

\newcommand{\RR}{\mathbb{R}}
\newcommand{\CC}{\mathbb{C}}
\newcommand{\minimize}{\mathop{\text{minimize\ }}}

\newcommand{\subjectto}{\mathop{\text{subject\ to\ }}}
\newcommand{\Ss}{\set{S}}
\newcommand{\vJ}{\boldsymbol{J}}

\newcommand{\vF}{\boldsymbol{F}}

\newcommand{\vr}{\boldsymbol{r}}

\newcommand{\hx}{\hat{\boldsymbol{x}}}

\newcommand{\hz}{\hat{\boldsymbol{z}}}

\newcommand{\hr}{\hat{\boldsymbol{r}}}
\newcommand{\hu}{\hat{\boldsymbol{u}}}

\newcommand{\vE}{\boldsymbol{E}}

\usepackage{xcolor}
\usepackage[normalem]{ulem}
\begin{document}

\title{Antenna Current Optimization and Realizations for Far-Field Pattern Shaping}

%% Use \affref{nn} and matching \aff{nn}{...} below for several authors
%% Mark the presenting author with an asterisk
\author{Shuai Shi,
Lei~Wang,
and B. L. G. Jonsson\\
 
\thanks{Manuscript received \today.}

\thanks{S. Shi and B.L.G. Jonsson are with Electromagnetic Engineering Lab of School of Electrical Engineering, KTH Royal Institute of Technology, SE-100 44 Stockholm, Sweden (e-mail: shuaishi@kth.se). L. Wang is with Institut f{\"u}r Theoretische Elektrotechnik, Hamburg University of Technology, Hamburg, Germany.

}% <-this % stops a space
}

\markboth{SHI+etal}
{Shi \MakeLowercase{\textit{et al. }}: Antenna Current Optimization and Realizations for Far-Field Pattern Shaping} % The only time the second header will appear is for the odd numbered pages

\maketitle
% * <a525154514@gmail.com> 2017-05-05T10:00:45.520Z:
%
% ^.
\begin{abstract} %IET max 200 words

Far-field shaping of small antennas is a challenge and the realizations of non-dipole radiation of small to intermediate sized antennas are difficult. Here we examine the antenna bandwidth cost associated with such constraints, and in certain cases we design antennas that approach the bounds. Far-field shaping is in particular interesting for Internet-of-things (IoT) and Wi-Fi applications since e.g. spatial filtering can mitigate package loss through a reduction of mutual interference, and hence increase the power efficiency of the devices.

Even a rather careful far-field shaping of smaller antennas can be associated with a steep reduction in the best available bandwidth.  It is thus important to develop constraints that a small antenna can support. We describe a power front-to-back ratio, and a related, beam-shaping constraint that can be used in optimization for the minimum Q-factor. We show that such a non-convex Q-factor optimization can be solved with the semi-definite relaxation technique.  
%The here presented results is an extension of the convex current optimization techniques, that successfully predict reachable bounds on antenna bandwidth. 
We furthermore show that certain of the above optimized non-standard radiation patterns can be realized with a multi-position feeding strategy with a moderate loss of Q-factor: $Q_\text{antenna}\leq 1.61Q_\text{optimal}$.

\end{abstract}


\section{Introduction}

\IEEEPARstart{S}{mall} antennas are receiving an increased attention in a number of application areas, including Internet-of-things (IoT), wearable antennas, applications with embedded antennas and low-frequency antennas. Common for the area is a scarcity of power and that a small bandwidth suffices. These devices are often driven from a battery or harvested energy. The targeted type of application often works on the ISM-band in Europe at 0.868 GHz with a 1\% bandwidth. In fact, a channel bandwidth of 1 MHz is often enough for certain IoT sensors, e.g., a sensor monitoring the temperature or to determine the location by transmitting the GPS information~\cite{Nokia2017}. In these kind of system it is known that loss and subsequent re-sending of a communication package due to e.g. interference is an essential factor in the overall energy consumption of the system~\cite{Pegatoquet2016}. Pattern shaping, spatial filtering through a high front-to-back ratio, or beam-shaping are possible tools to mitigate the interference in these kind of communication system. Similarly for wearable devices, polarization and pattern shape can strongly influence the efficiency and feasibility of such devices. Such far-field shaping tends to come with a cost in bandwidth which is the topic of this paper.

In this paper we develop new tools to investigate bounds on the best possible bandwidth-performance under different far-field constraints. We also consider a multi-position feeding strategy inspired by the optimal current. This strategy is used to realize a  selection of  antennas that approach the bounds. The size of the considered antenna is equal to or smaller than the electrical size of a dipole antenna. The new far-field constraints that we develop include a power version of the front-to-back ratio (FBR) and a related `beam-shaping' constraint. 

We begin with an investigation of the effects of adding such non-convex far-field constraints to a Q-factor optimization. As illustrated here these non-convex problem can be solved deterministically, without resorting to global genetic-like tools.
We determine the bandwidth associated cost of such far-field shaping for certain small antennas. Convex far-field constrains and their relation to the antenna bandwidth has been considered earlier in e.g.~\cite{Gustafsson+Nordebo2013,Gustafsson+etal2016a}. We have also investigated the non-convex constraints with regard to super-directivity in
~\cite{Jonsson+etal2017b,Fabien+etal2017}. It is well known that super-directivity is associated with a narrow bandwidth see e.g.~\cite{Hansen2006,Yaghjian+etal2008,Gustafsson+Nordebo2013}. 

There are different approaches to find the physical bounds of antennas, e.g., circuit models, vector models, sum rules and stored energy~\cite{Chu1948,Thal1978,Hansen1981,Collin+Rothschild1964,Geyi2003,Vandenbosch2010,Gustafsson+etal2012a,Gustafsson+Nordebo2013,Jonsson+Gustafsson2015,Capek+Jelinek2016,Schab+etal2017}. The Q-factor approach~\cite{Vandenbosch2010,Gustafsson+etal2012a,Gustafsson+Nordebo2013,Jonsson+Gustafsson2015,Capek+Jelinek2016,Schab+etal2017} for small antennas is a good approximation for the bandwidth~\cite{Yaghjian+Best2005,Gustafsson+Jonsson2015a}. It furthermore provides interesting and useful information about antenna design possibilities~\cite{Yaghjian+Best2005,Gustafsson+Nordebo2013,Jonsson+Gustafsson2015,Capek+Jelinek2016,Schab+etal2017}. The first limitation in terms of Q-factor for small antennas bounded by sphere was given by Wheeler and by Chu in 1940s~\cite{Wheeler1947,Chu1948}. Recently the Q-factor limitations on small antennas have been developed to tighter bounds for arbitrarily shaped antennas. These bounds can be used to predict optimal bandwidth with certain far- and near-field constraints as well as partial directivity constraints~\cite{Gustafsson+Nordebo2013}. These problems were shown to be convex, which makes them easy to solve deterministically. However there are also a number of constraints which cannot be treated with the existing fast methods. These problems include total directivity, radiated power, Q-factors, etc. Recently, an eigenvalue based method has been developed for certain non-convex constraints, see e.g.~\cite{Jelinek+Capek2017,Jonsson+etal2017b}.

With the observation~\cite{Jonsson+etal2017b} that the semi-definite relaxation (SDR) technique~\cite{Lovasz1979,Goemans+Williamson1995,Luo+etal2007,Luo+etal2010} applies to a this challenging class of antenna optimization problems, we have obtained a new tool to investigate how the bandwidth depends on quadratic antenna constraints. In this paper, we illustrate that power-pattern constraints can be determined deterministically using CVX~\cite{Boyd2004,Grant+Boyd2014,Vandenberghe+Boyd1996}. They include power radiation pattern shaping and Q-factor. To obtain these bounds we utilize stored energies, which has been extensively examined in~\cite{Yaghjian+Best2005,Gustafsson+Jonsson2015a,Geyi2015,Vandenbosch2010,Gustafsson+Nordebo2006,Collin+Rothschild1964,Fante1969,Pozar2009,Gustafsson+etal2012a,Capek+Jelinek2015b,Jonsson+Gustafsson2015,Capek+etal2013,Geyi2003,Stuart+Best+Yaghjian2007,Gustafsson2009}. Since these power-associated constraints appear in quadratic forms, they are included in the class of quadratically constrained quadratic programs (QCQPs) see e.g ~\cite{Boyd2004}, and it is known that the non-convex QCQP problems can be general NP-hard. For NP-hard problems, there is no efficient algorithms yet that can find global optimal solutions, for a recent discussion  see e.g.~\cite{Boyd2017}. For some special cases with few constraints, we can use the SDR technique to get accurate approximations, or under specific conditions, the exact optimal solutions~\cite{Luo+etal2007,Luo+etal2010,Shi+etal2017,Jonsson+etal2017b,Jonsson+etal2017}. In recent years, SDR technique has been known as an efficient high-performance approach in the area of signal processing and communications, for instance, for MIMO detection, sensor network localization, and phase retrieval~\cite{Luo+etal2010,Candes+etal2015,Luo+etal2007}. It is also used to find the physical limitation of MIMO antennas, and the upper bound of 5G RF electromagnetic field exposure~\cite{Ehrenborg+Gustafsson2017,Xu2017}, and in wireless power transfer~\cite{Lang2017}.

It is very interesting to investigate and design optimal antennas which are close to the physical limitations. The performance properties of several fundamental small antenna designs, e.g., the folded spherical and cylindrical helices, were examined and compared with their physical limitations in~\cite{Best2007}.Yaghjian designed an electrically small two-element super-directivity array with near optimal end-fire directivity in~\cite{Yaghjian+etal2008}. A two-element parasitic antenna with high directivity and low Q-factor was designed and compared with its bound in~\cite{Fabien+etal2017}, see also~\cite{Capek2018}. The idea of using multi-position feedings to approximate optimal current was proposed in~\cite{Jelinek+Capek2017}. This can lead to a suboptimal solution for antenna design, however we show that multi-position feedings can approach the bound. Here in this paper we use a multi-position feeding strategy to realize desired radiation patterns for dipole antennas, and compare the Q-factors with the corresponding limitations.

The organization of this paper is as follows. In Section II, we introduce the Q-factor for small antennas, state the constrained minimization problems for the Q-factor, and introduce the SDR technique required to rapidly solve these minimization problems. In Section III, the SDR technique is applied to antenna current optimization problems of minimizing the stored energy with power-based radiation constraints. In particular a sectoral beamwidth and a power front-to-back ratio are investigated, and the corresponding lower bounds on the Q-factor are determined. In Section IV, we consider the optimal Q-factor for a given directivity for a small array. In Section V, we use multi-position feedings to realize novel radiation patterns for the dipole. The paper ends with conclusions and a bibliography.

\section{Optimization Problems}

\subsection{Q-factor}

The fractional impedance bandwidth (BW) of an antenna at a given maximal allowed reflection coefficient $|\Gamma_0|$ is often a critical parameter in antenna designs. It was shown in~\cite{Yaghjian+Best2005} that for small antennas we have the relation 
\begin{equation}\label{BW}
BW = \frac{f_2-f_1}{f_0}\approx \frac{2|\Gamma_0|}{Q_Z\sqrt{1-|\Gamma_0|^2}},
\end{equation}
see also~\cite{Gustafsson+Nordebo2006}. This relation connects the fractional bandwidth expressed as the ratio of the frequency band $f_2-f_1$ to the the center frequency $f_0=(f_1+f_2)/2$. Here $Q_Z$ is the input impedance based estimation of the Q-factor. Given the antenna port impedance 
\begin{equation}\label{Zin}
Z_{\text{in}}=R_\text{in}+\ju X_\text{in},%=\frac{2\rP+4j\omega(\Wm-\We)}{|I_\text{in}|^2}.
\end{equation}
we have 
\begin{equation}\label{Qz}
Q_Z = \frac{\sqrt{(\omega R'_\text{in})^2+(\omega X'_\text{in}+|X_\text{in}|)^2}}{2R_\text{in}},
\end{equation}
where $R_\text{in}$ and $X_\text{in}$ are the real and imagery part of the input impedance $Z_{\text{in}}$ respectively, and $'$ denotes the derivative with respect to angular frequency. It has been shown in~\cite{Yaghjian+Best2005,Gustafsson+Jonsson2015a} that the Q-factor is a good estimation of the bandwidth for narrow band antenna. Utilizing~\eqref{Qz} we can determine $Q_Z$ when we have a realized antenna. 

To determine the best physical bound among all possible antennas with respect to bandwidth that fit into a given geometry we need a slightly different tool than $Q_Z$. The tool that we use is a Q-factor that is based on stored energies, see e.g.~\cite{Vandenbosch2010,Gustafsson+etal2012a}. It is known that for well designed antennas with large enough Q-factor ($Q\geq 5$),  we have $Q\approx Q_Z$, see e.g.~\cite{Gustafsson+Jonsson2015a}, thus to maximize the fractional bandwidth it suffices to minimize the Q-factor. For a lossy antenna, it is easy to increase the Q-factor by increasing losses. It is thus harder to obtain a good bandwidth for a lossless antenna, and we will in this paper restrict our considerations to lossless antennas.

The lossless Q-factor based on stored energy is defined as ratio between the time-average stored energy and the radiated power, see e.g.~\cite{Gustafsson+Jonsson2015a}
\begin{equation}\label{Qdef}
Q = \frac{2\omega\max(\We,\Wm)}{\rP},
\end{equation}
where $\omega$ is the angular frequency, $\We$ and $\Wm$ are the stored electric and magnetic energy respectively. The total radiated power $\rP$ of the antenna is defined as
\begin{equation}\label{Prad}
\rP = \int_{\Ss^2} U(\hr)\diff \Omega = \frac{1}{2\eta} \lim_{R_0\rightarrow \infty }\int_{|\vr|=R_0} |\vE(\vr)|^2\diff S,
\end{equation}
where $\Ss^2$ is the unit sphere in $\RR^3$, $U(\hr)$ is the radiation intensity in the direction $\hr$, $\diff \Omega=\sin\theta\diff \theta\diff \phi$, where we let $\theta$ be the polar angle and $\phi$ be the azimuth angle. Here $\vE(\vr)$ is the radiated electric field at the point $\vr$, $\diff S=r^2\diff \Omega$, and $\eta$ is the free space impedance. In this paper, we use \eqref{Qdef} to calculate the value of the Q-factor for antennas.

\subsection{Current representation}

For lossless metal antennas, all the above discussed antenna parameters can be inferred from the surface current density $\vJ$, that are localized on a surface $S$. Antenna current optimization gives a physical limitation and an a priori estimate for all possible designs of small passive antennas in a given region enclosed by $S$, see e.g.~\cite{Gustafsson+Nordebo2013,Gustafsson+etal2016a}. The underlying idea is that optimization over all surface currents on $S$ includes the current of the best antenna within $S$. The stored electric and magnetic energies for surface currents are given by e.g.~\cite{Vandenbosch2010,Jonsson+Gustafsson2015,Gustafsson+etal2016a}:
\begin{multline}\label{WeJ}
\We = \frac{\eta}{4\omega}\int_S\int_{S} (\nabla_1\cdot \vJ_{1})(\nabla_2\cdot\vJ_{2}^*)\frac{\cos(kr_{12})}{4\pi kr_{12}}\\
-\big(k^2\vJ_{1}\cdot\vJ_{2}^ *-(\nabla_1\cdot\vJ_{1})\nabla_2\cdot \vJ_{2}^*\big)\frac{\sin(kr_{12})}{8\pi}
\diff S_1\diff S_2,
\end{multline}
and
\begin{multline}\label{WmJ}
\Wm = \frac{\eta}{4\omega}\int_S\int_{S} k^2\vJ_{1}\cdot\vJ_{2}^*\frac{\cos(kr_{12})}{4\pi kr_{12}}\\
-\big(k^2\vJ_{1}\cdot\vJ_{2}^ *-(\nabla_1\cdot\vJ_{1})\nabla_2\cdot \vJ_{2}^*\big)\frac{\sin(kr_{12})}{8\pi}
\diff S_1\diff S_2,
\end{multline}
where we use the notation $\vr_1,\vr_2\in \RR^3$, $\vJ_{1}=\vJ(\vr_1)$, $\vJ_{2}=\vJ(\vr_2)$, $r_{12}=|\vr_1-\vr_2|$. 
%, and $\eta$ is the free space wave impedance. (see above)
Similarly, the total radiated power is
\begin{multline}\label{PradJ}
\rP = \frac{\eta}{2}\int_S\int_{S}( k^2\vJ_{1}\cdot\vJ_{2}^*\\
-(\nabla_1\cdot\vJ_{1})\nabla_2\cdot \vJ_{2}^*)\frac{\sin(kr_{12})}{4\pi kr_{12}}
\diff S_1\diff S_2.
\end{multline}
We can also express the far field radiation pattern using surface current density $\vJ$. The far field radiation intensity is 
\begin{equation}\label{UJ}
U(\hr) = \frac{\eta}{32\pi^2}\int_S\int_{S} k^2\vJ_{1}\cdot\vJ_{2}^*e^{jk(\vr_1-\vr_2)\cdot\hr}\diff S_1\diff S_2.
\end{equation}
 
\subsection{MoM approach}

To numerically determine the stored energy and radiated power, we use an in-house Method of Moment (MoM) approach based on the Rao-Wilton-Glisson (RWG) basis functions $\boldsymbol{\psi}_n(\vr)$~\cite{RWG1982}. The surface current density is approximated by
\begin{equation}\label{base}
\vJ(\vr) \approx \sum_{n=1}^{N} I_n \boldsymbol{\psi}_n(\vr),
\end{equation}
where we introduce an N$\times$1 current vector $\mathbf{I}$ with elements $I_n$ to simplify the notation. Noticing that $\We$, $\Wm$ largely correspond to components in the electric field integral used in MoM, we find the matrix formulations of the stored energy and the radiated power as well as the power intensity, following the notation of~\cite{Gustafsson+etal2016a}:
\begin{equation}\label{WeI}
\We \approx \frac{1}{8}\mathbf{I^H}\big(\frac{\partial\mathbf{X}}{\partial\omega}-\frac{\mathbf{X}}{\omega}\big)\mathbf{I}=\frac{1}{4\omega}\mathbf{I^H}\mathbf{X_\mathrm{e}}\mathbf{I},
\end{equation}
\begin{equation}\label{WmI}
\Wm \approx \frac{1}{8}\mathbf{I^H}\big(\frac{\partial\mathbf{X}}{\partial\omega}+\frac{\mathbf{X}}{\omega}\big)\mathbf{I}=\frac{1}{4\omega}\mathbf{I^H}\mathbf{X_\mathrm{m}}\mathbf{I},
\end{equation}
\begin{equation}\label{PradI}
\rP\approx \frac{1}{2}\mathbf{I^H}\mathbf{R}\mathbf{I},
\end{equation}
\begin{equation}\label{UI}
U(\hr)\approx \frac{1}{2\eta}|\mathbf{FI}|^2=\frac{1}{2\eta}\mathbf{I^H}\big(\mathbf{F^HF}\big)\mathbf{I},
\end{equation}
\begin{equation}\label{PI}
P_{\Omega_0} \approx\frac{1}{2\eta}\mathbf{I^H}\big(\int_{\Omega_0} \mathbf{F^HF} \diff \Omega\big) \mathbf{I},
\end{equation}
where $\mathbf{R}=\mathop{\rm Re}{(\mathbf{Z})}$, $\mathbf{X}=\mathop{\rm Im}{(\mathbf{Z})}$, and $\mathbf{Z}$ is the MoM impedance matrix. 

The far-field $\vF(\hr)$ is defined from the electric field by $r\vE(\vr)\lexp{\ju kr} \rightarrow \vF(\hr)$ as $r\rightarrow \infty$, and it is approximated by the vector $\mathbf{F}(\hr)$ through $\vF(\hr)\approx \mathbf{F(\hr)I}$. Thus we have that $U(\hr)=|\vF(\hr)|^2/(2\eta)$, which agrees with \eqref{UI}. $P_{\Omega_0}$ denotes the radiated power in the region $\Omega_0\subset\mathrm{S}^2$. It is clear that all the above antenna quantities (except $\mathbf{F}$) are quadric forms of the current $\mathbf{I}$.

\subsection{Semi-definite relaxation}

Convex optimization has sucessfully been used to determine physical bounds on the Q-factor under certain convex constraints. 
In this paper we extend the class of convex constraints to a larger class of generic quadratic constraints. All the investigated problems in this paper can be reduced to the following general class of minimizing problems:
\begin{equation}\tag{Q}\label{MinQ}\begin{aligned}
\minimize_{\mathbf{I}\in \CC^N} & \max\{\mathbf{I^H}\mathbf{X_\mathrm{e}}\mathbf{I},\mathbf{I^H}\mathbf{X_\mathrm{m}}\mathbf{I}\big\}\\
\subjectto & \mathbf{I^H}\mathbf{M}_i\mathbf{I}=a_i,\\
& \mathbf{I^H}\mathbf{N}_j\mathbf{I}\geq b_j.
\end{aligned}\end{equation}
where $\CC^N$ is $N$-vectors with complex coefficients, $\mathbf{M}_i$ and $\mathbf{N}_j$ are $N \times N$ matrices, $a_i$ and $b_j$ are constants, $i=1, 2, ..., \mathcal{M}$, $j=1, 2, ..., \mathcal{N}$ indicates the number of different constraints. With different $\mathbf{M_\text{i}}$ and $\mathbf{N_\text{j}}$, we can formulate antenna current optimization problems, to determine the Q-factor with constraints on e.g. directivity, radiation patterns, etc.

The problem (Q) is here in general non-convex for matrices $\mathbf{M}_i$ and $\mathbf{N}_j$. Certain non-convex problem can be relaxed into a convex problem, see e.g.~\cite{Boyd2004}. Here we apply a technique called semi-definite relaxation (SDR) to relax (Q) into a convex, semi-definite problem~\cite{Luo+etal2010,Lovasz1979,Goemans+Williamson1995,Luo+etal2007,Boyd2017}, see also~\cite{Shi+etal2017,Jonsson+etal2017,Jonsson+etal2017b}. The procedure is as follows:
\begin{equation}
\mathbf{I^HM_\text{i}I}=\mathop{\rm tr}{(\mathbf{I^HM_\text{i}I})}=\mathop{\rm tr}{(\mathbf{M_\text{i}II^H})},
\end{equation}
and similarly for $\mathbf{N_\text{j}}$, $\mathbf{X_\mathrm{e}}$, $\mathbf{X_\mathrm{m}}$. We can thus rewrite (Q) as
\begin{equation}\tag{R}\label{pQc}\begin{aligned}
\minimize_{\mathbf{K}\in \CC^{N\times N}, \mathbf{K}\geq\mathbf{0}} & \max\{\mathop{\rm tr}{(\mathbf{X_\mathrm{e}K})},\mathop{\rm tr}{(\mathbf{X_\mathrm{m}K})}\big\}\\
\subjectto & \mathop{\rm tr}{(\mathbf{M}_i\mathbf{K})}=a_i,\\
& \mathop{\rm tr}{(\mathbf{N}_j \mathbf{K})}\geq b_j.
\end{aligned}\end{equation}
where $\mathbf{K}= \mathbf{II^H}$. The problem (R) together with $\rank(\mathbf{K})=1$ is equivalent with (Q)~\cite{Luo+etal2010,Lovasz1979,Goemans+Williamson1995,Luo+etal2007}. We observe that the constraints in (R) are affine functions in the matrix $\mathbf{K}$, and that $\mathbf{K}= \mathbf{II^H}$ is equivalent to that $\mathbf{K}$ is a rank one semidefinite Hermitian matrix. If we drop the constraint $\rank(\mathbf{K})=1$, we get the relaxed problem (R), which is known as an SDR problem of (Q). (R) can be solved by many efficient methods, see e.g. \cite{ODonoghue+etal2016}. 

We can thus get a globally optimal solution $\mathbf{K_*}$ to problem (R), that clearly is a lower bound for the problem (Q). If $\mathbf{K_*}$ is of rank one, we determine $\mathbf{I_*}$ from $\mathbf{K}= \mathbf{I_*I_*^H}$, and $\mathbf{I_*}$ will be a feasible and optimal solution to the original problem. It is shown in~\cite{Huang+Zhang2007}, see also~\cite{Jonsson+etal2017b}, that if the number of the constraints $\mathcal{M}+\mathcal{N}$ is less or equal than 2, then there exist a solution with $\rank(\mathbf{K_*})=1$. If $\rank(\mathbf{K_*})> 1$, we need to convert $\mathbf{K_*}$ into a feasible approximation $\tilde{\mathbf{I}}$ to problem (Q) using the methods from e.g.~\cite{Luo+etal2010,Luo+etal2007}.

Above we have expressed stored energy, radiated power, etc., as quadratic forms of the antenna surface current density. The RWG-basis representation thus gives us a numerical estimate of the physical bound. Using the SDR technique we can investigate the limits of the best possible Q-factor in the antenna under a range of constraints on the far-field. 

\section{Power Pattern Shaping}

Design of large antennas frequently come with far-field requirements, often denoted pattern-shaping, see e.g.~\cite{Stutzman2012,Mohal2010,Bucci1994}. It is interesting to inquire whether similar constraints can be applied to small antennas, and what cost such requirements impose on the bandwidth. Clearly pattern-shaping of small antennas requires in general a weaker type of constraints than what large antennas can support. One such requirement is to consider the ratio of radiated power in the angular region $\Omega_0$ as compared with the total radiated power:
\begin{equation}\label{alpha}
\frac{P_{\Omega_0}}{\rP} \approx \frac{1}{\eta}\frac{\mathbf{I^H}\big(\int_{\Omega_0} \mathbf{F^HF} \diff \Omega\big) \mathbf{I}}{\mathbf{I^H}\mathbf{R}\mathbf{I}}=\alpha.
\end{equation}
Clearly $\alpha\in\lbrack0,1\rbrack$. 

To minimize the Q-factor with \eqref{alpha} as a far-field constraint, we formulate the optimization problem:
\begin{equation}\label{Shaping}\begin{aligned}
\minimize_{\mathbf{I}\in \CC^N} & \max\{\mathbf{I^H}\mathbf{X_\mathrm{e}}\mathbf{I},\mathbf{I^H}\mathbf{X_\mathrm{m}}\mathbf{I}\big\}\\
\subjectto & \mathop{\mathbf{I^H}\mathbf{R}\mathbf{I}}=1,\\
& \mathop{\mathbf{I^H}\big(\int_{\Omega_0} \mathbf{F^HF} \diff \Omega\big) \mathbf{I}}\geq\eta\alpha.
\end{aligned}\end{equation}
We call this type of minimization problems for far-field power pattern shaping. We use below this type of constraints to determine a kind of sector ``beamwidth'' of the pattern by requiring that $\alpha$ is high over a given sector. By utilizing the relation between the total radiated power and the we can also use this type of constraint to determine the band-width cost of having a `null' in the complementary region to $\Omega_0$, e.g. 
$\Omega_0=\mathrm{S}^2\backslash \{(\theta,\phi):|\theta-\theta_0|\leq \delta_0,|\phi-\phi_0|\leq\delta_0\}$ and $\alpha=1-\delta_1$, where $\delta_0,\delta_1$ are small. Convex far-field constraints for one polarization has previously been investigated in \cite{Gustafsson+Nordebo2013}, and the problem~\eqref{Shaping} thus extend this class of problems.

\subsection{Beamwidth}\label{sec.Beam}

\begin{figure}[htbp]
  \centering
  \includegraphics[width=\linewidth]{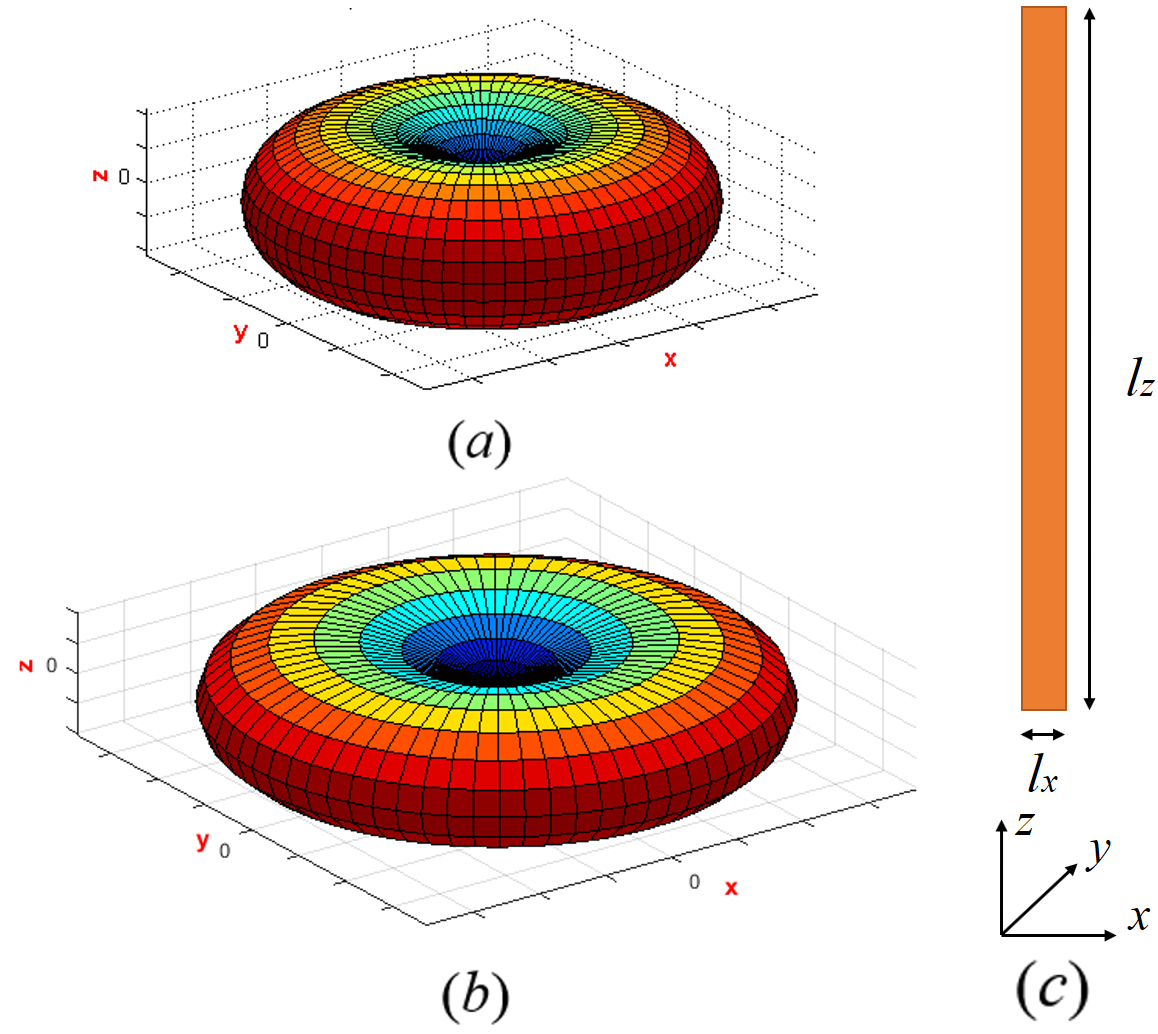}
   \caption{3D radiation patterns. (a) The reference is the unconstrained radiation pattern of the dipole.  (b) The radiation pattern associated with the optimal Q-factor solving \eqref{Shaping} with $\Omega_1$ and $\alpha=0.5$. }
  \label{flat3D}
\end{figure}

We utilize the power pattern constrained Q-factor problem~\eqref{Shaping} to determine the minimum Q-factor with constraints on the amount of power in a given angular sector. Such a constraint results in increased/reduced sector-beamwidth. We consider three sectors $\Omega_\kappa$ are sectors, $\kappa=1,2,3$ where 
\begin{equation}\label{O1}
\Omega_1=\{\vr\in\RR^3:75^\circ<\theta<105^\circ\},
\end{equation}
\begin{equation}
\Omega_2=\{\vr\in\RR^3:70^\circ<\theta<110^\circ\},
\end{equation}
\begin{equation}\label{O3}
\Omega_3=\{\vr\in\RR^3:60^\circ<\theta<120^\circ\}.
\end{equation}
$\Omega_1$ to $\Omega_3$ describe subsequently wider sectors around the main direction of a classical dipole pattern. The optimization problem~\eqref{Shaping} is clearly applicable to any arbitrary antenna shape, that is sufficiently small so that the stored energy remains positive~\cite{Gustafsson+etal2012a}. Here we investigate how the strip dipole, see Fig. \ref{flat3D} reacts on increasing demands of a narrower radiation direction. The reason that we consider the dipole is mainly due to the realization strategy as discussed in Section \ref{sec:Design}. The strip dipole is infinitely thin, perfect electric conducting (PEC) and has length $l_z$ and width $l_x=0.02l_z$. We choose the frequency such that $l_z=0.48\lambda$, or $kl_z=0.48\cdot2\pi$, where where $k$ is the wave number. Here we choose the frequency to 0.868 GHz which is for the ISM applications in Europe. To numerically investigate this case, a mesh with $N_x=2$ in $x$-direction and $N_z=100$ in $z$-direction is used. A finer meshes tend to result in slightly smaller Q-factor.

\begin{figure}[htbp]
  \centering
  \includegraphics[width=\linewidth]{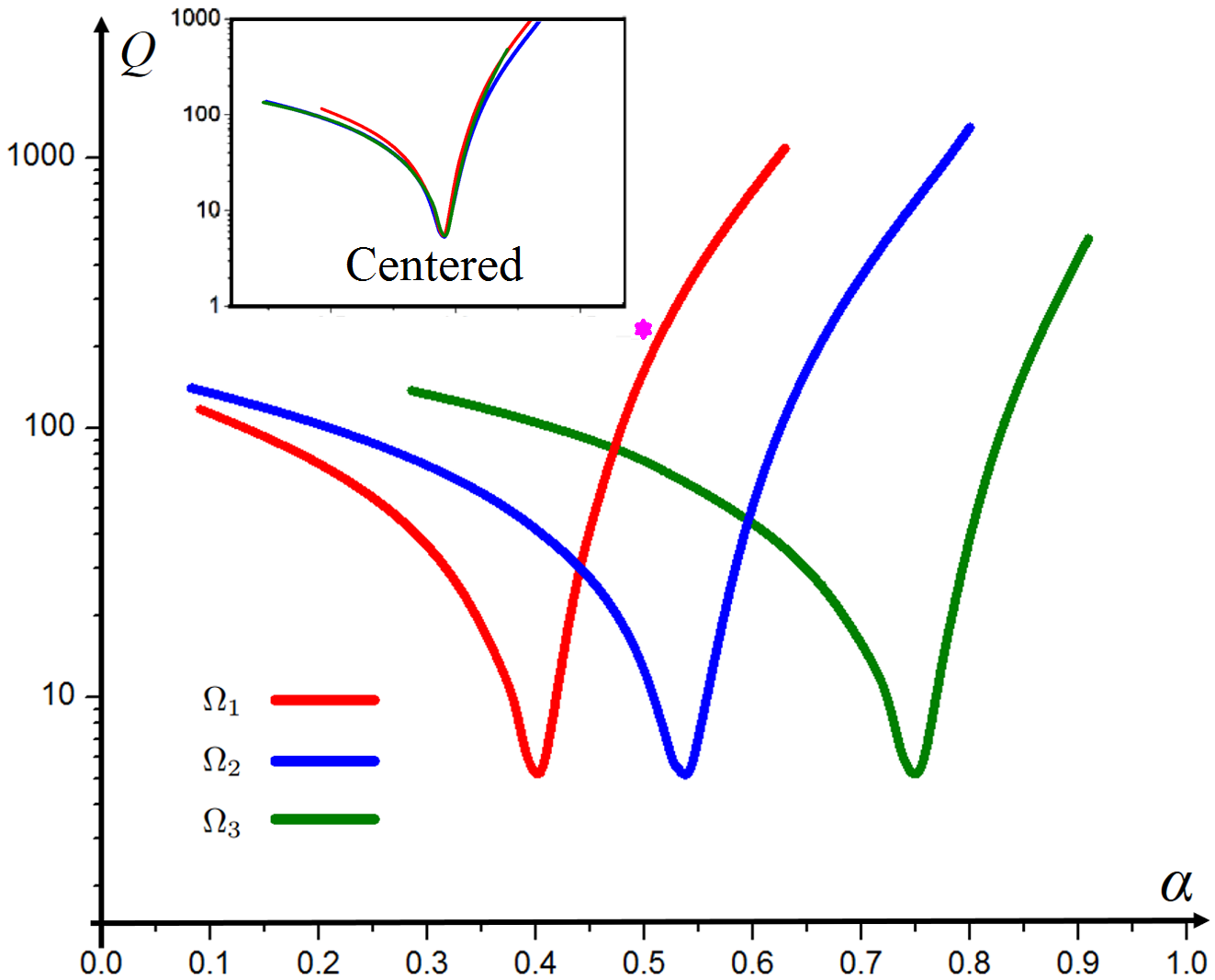}
   \caption{The lower bounds on the Q-factor for a given $\Omega$ and $\alpha$ for the strip dipole. }
  \label{flatQ}
\end{figure}

\begin{figure}[htbp]
  \centering
  \includegraphics[width=0.8\linewidth]{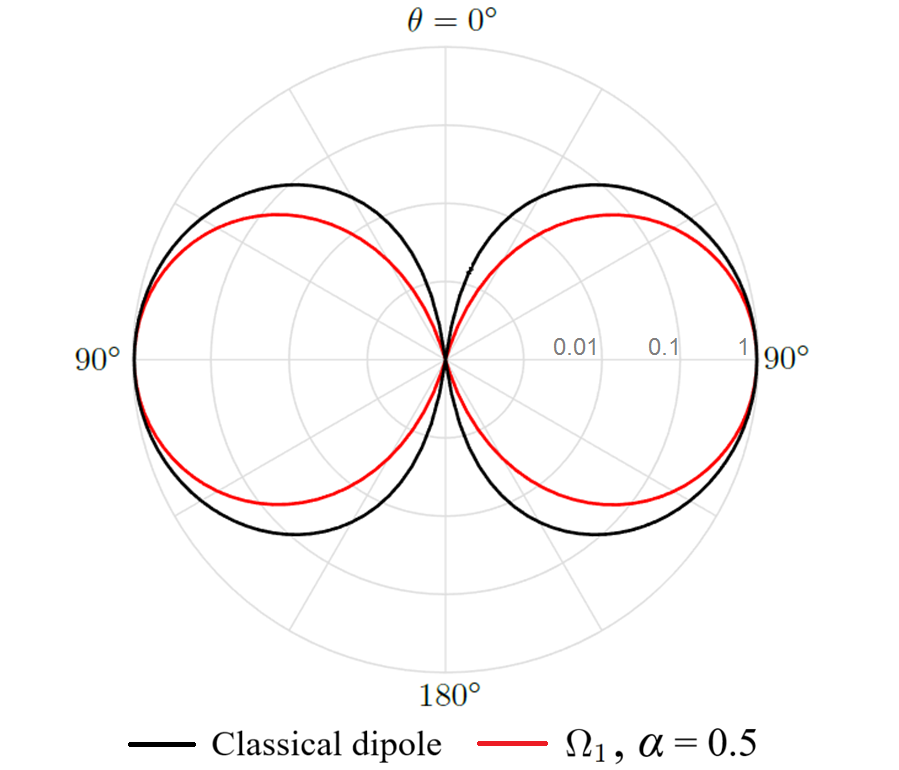}
   \caption{The 2D radiation patterns (normalized in logarithmic scale) with different beamwidth for the strip dipole, $xz$-plane.}
  \label{flat2D}
\end{figure}

The optimal Q as a function of $\alpha$ is depicted in Fig. \ref{flatQ}. Here we sweep the parameter $\alpha$ for different choices of the size of the sectors $\Omega_\kappa$. Clearly, for certain $\alpha$-values the constraint corresponds to an unperturbed dipole and we get the lowest Q-factor. These values are at $\alpha=[0.4, 0.53, 0.74]$, for $\Omega_\kappa$, $\kappa=1,2,3$ respectively. The associated Q-factor at this minimum point is close to the result in~\cite{Jonsson+Gustafsson2015}:
\begin{equation}
Q=\frac{6\pi}{k^3\bm{\gamma}_e},
\end{equation}
valid for $ka\ll 1$. Here $\bm{\gamma}_e$ is the largest eigenvalue of the electric polarizability matrix, and $a$ is the minimum radius of a sphere enclosing the antenna. 

Demands of more or less power in the region $\Omega_\kappa$ are associated with a cost of rapidly decreasing bandwidth associated with a higher Q-factor. It is interesting to observe that it is less bandwidth-expensive to reduce amount of power in the $\Omega_\kappa$ region than to increase it, as is clear from the asymmetry of the graphs in Fig. \ref{flatQ}. In the inset in Fig. \ref{flatQ}, where we shift the three curves so that the lowest points are centered, we see the expected same minimum Q-factor value associated with a dipole and the similarity of the curves. An associated 
radiation pattern for $\Omega_1$ and $\alpha=0.5$ together with the reference dipole are shown both in 3D in Fig. \ref{flat3D}, and in 2D in Fig. \ref{flat2D}. 
The patterns are normalized such that the total radiated power is one. Realization of an antenna with this type of ``beam shaping'' is also discussed in Section \ref{sec:Design}.

The oblique pattern shaping imposed with \eqref{O1}--\eqref{O3} are here fitted to that we only have a small region where the surface current can be. A wider antenna is expected to have less increase in best Q-factor under these constraints, due to its higher polarizability.

\subsection{Power front-to-back ratio}\label{sec.PFBR}

The front-to-back ratio (FBR) is an often used measure in beam-type antennas, see e.g.~\cite{IEEEstandard}. However, recently FBR has also been suggested as a first step of spatial filtering~\cite{Jonsson+etal2017b,Fabien+etal2017}. Spatial filtering can be used to mitigate package loss in communication systems by reducing interference~\cite{Pegatoquet2016} and hence reducing the overall energy consumption of the system. There exist several definitions of FBR. Max/min comparison in one direction is considered in the standard of IEEE~\cite{IEEEstandard}. It is also widely used by many companies, e.g. see~\cite{Cisco2007}. Another definition by NGMN Alliance~\cite{NGMN2013} uses the radiation in the rear $\pm 30 ^{\circ}$ angular region, instead of a single rear direction with the forward direction, to calculate the FBR. 

Here we define a power front-to-back ratio (PFBR), where we investigate the FTB-ration not in just opposite directions, but as a ratio of the radiated power for two different regions of the sphere. This kind of PFBR is more suitable for small antennas, and with an appropriate choice of regions it includes the above two definitions. In the present case we are concerned with the total radiated power, but modification to one polarization is straight forward.

Consider two non-overlapping regions $\Omega_+$ and $\Omega_-$, we will call $\Omega_+$ the forward region and $\Omega_-$ is the backward region. 
The PFBR is defined as 
\begin{equation}
\text{PFBR}=\frac{\int_{\Omega_+}U(\hr)\diff \Omega}{\int_{\Omega_-}U(\hr)\diff \Omega}.
\end{equation}
Thus the quantity PFBR defines the ratio of the radiated power of the $\Omega_+$-region to the $\Omega_-$-region. 
A larger PFBR suppresses radiation in $\Omega_-$, while increasing it in $\Omega_+$. 
Rewriting the constraint in the form 
\begin{equation}
 \mathop{\mathbf{I^H}\big(\int_{\Omega_+} \mathbf{F^HF} \diff \Omega\big) \mathbf{I}}  - \text{PFBR }\mathop{\mathbf{I^H}\big(\int_{\Omega_-} \mathbf{F^HF} \diff \Omega\big) \mathbf{I}}\geq 0,
\end{equation}
illustrates that the constraint is compatible with the problem (Q), and it can hence be reduced by the SDR-method. 

Utilizing that $\int_{\mathrm{S}^2} |\mathbf{FI}|^2 \diff \Omega$ is proportional to the total radiated power we see that with an appropriate choice of $\Omega_\pm$ in the PFBR definition generalizes the pattern-shaping constraint in~\eqref{Shaping}. However, for other choices of $\Omega_\pm$ we have a more general constraint. 

For the present case we let $\Omega_\pm$ be the respective half-sphere defined by 
\begin{equation}\label{pQa}\begin{aligned}
\Omega_+(\hu)=\{\hr\in\RR^3:\hu\cdot\hr>0 \},
\end{aligned}\end{equation}
\begin{equation}\label{pQb}\begin{aligned}
\Omega_-(\hu)=\{\hr\in\RR^3:\hu\cdot\hr<0 \},
\end{aligned}\end{equation}
where $\hr$ is the observation direction, and $\hu$ indicates the `forward' direction.

We consider the optimal Q-factor problem with the PFBR constraint, see also~\cite{Shi+etal2017}:
\begin{equation}\label{PFBR}\begin{aligned}
\minimize_{\mathbf{I}\in \CC^N} & \max\{\mathbf{I^H}\mathbf{X_\mathrm{e}}\mathbf{I},\mathbf{I^H}\mathbf{X_\mathrm{m}}\mathbf{I}\big\}\\
\subjectto & \mathop{\mathbf{I^H}\mathbf{R}\mathbf{I}}=1,\\
& \frac{\mathop{\mathbf{I^H}\big(\int_{\Omega_+} \mathbf{F^HF} \diff \Omega\big) \mathbf{I}}}{\mathop{\mathbf{I^H}\big(\int_{\Omega_-} \mathbf{F^HF} \diff \Omega\big) \mathbf{I}}}\geq\text{PFBR}.
\end{aligned}\end{equation}

To determine the minimal Q-factor of~\eqref{PFBR} we consider two different constrain choices of $\Omega_\pm$: $\Omega_\pm(\hz)$ and $\Omega_\pm(\hx)$. 
We start with the constraints applied on a dipole, with applications in Sec. \ref{sec:Design}, and later on of a sweep of a rectangular shape. We express the result in terms of the fractional bandwidth utilizing the approximation \eqref{BW}: $BW\approx 2/Q$, for a reflection factor $|\Gamma|\leq 1/\sqrt{2}$. 

\begin{figure}[htbp]
  \centering
  \includegraphics[width=\linewidth]{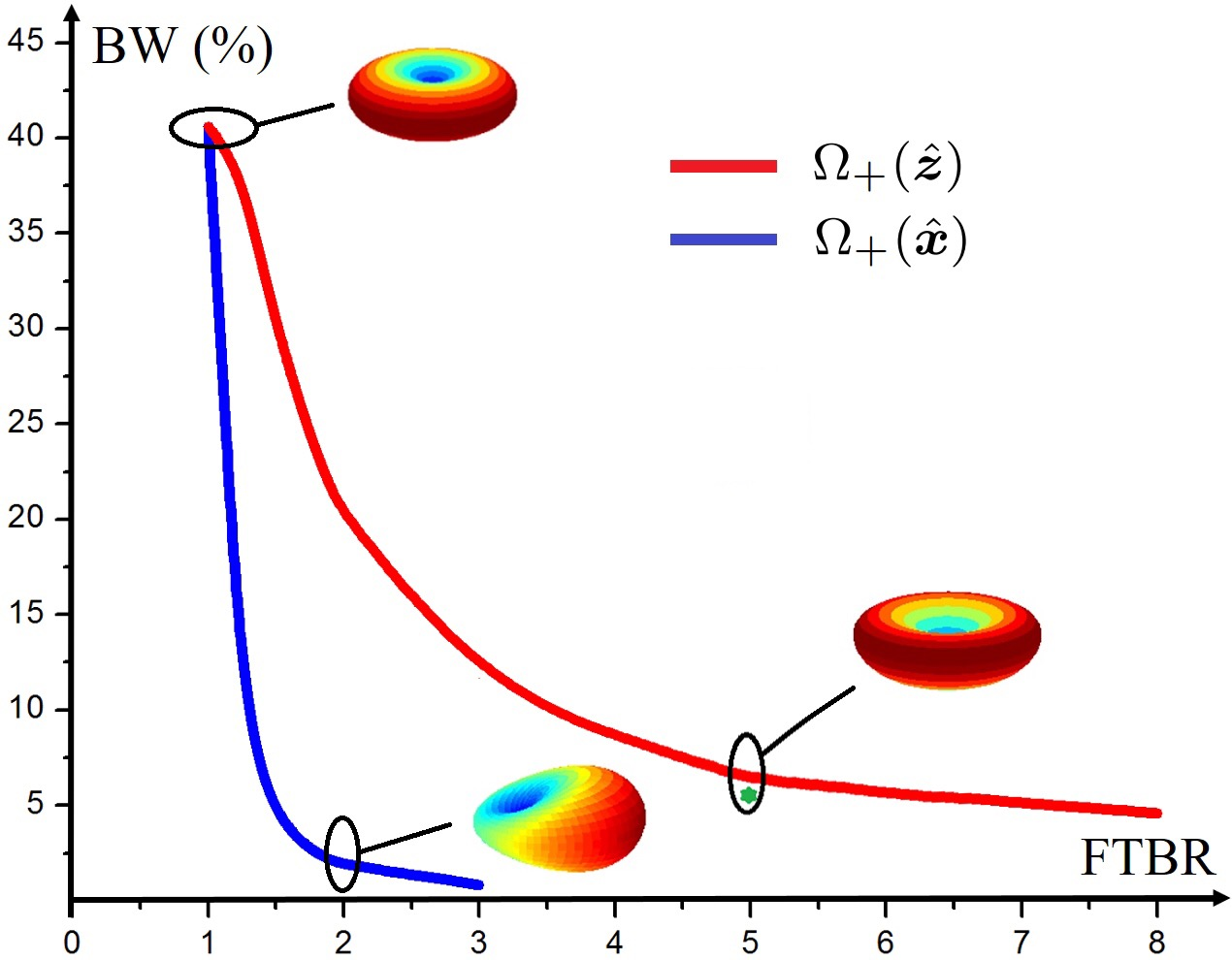}
   \caption{The bounds on the fractional bandwidth BW (\%) for a given PFBR for the strip dipole, for $|\Gamma_0|=1/\sqrt{2}$.}
  \label{fbrQ}
\end{figure}

In Fig.~\ref{fbrQ} we depict the 
maximal bounds on the fractional bandwidth for a dipole for a given PFBR. We include both $\hu=\hz$ and $\hu=\hx$ for completeness, as as expected PFBR with $\Omega_\pm(\hx)$ have a steep indeed decrease in Q-factor with increasing PFBR. The reason is clear, there is no space for the current to shape a pattern in this direction. For the $\Omega_\pm(\hz)$ we get a more interesting case with a tilted-radiation pattern. We note that a dipole shape can support a PFBR of 5 in $\Omega_+(\hz)$ with about 8\% fractional bandwidth.

\begin{figure}[htbp]
  \centering
  \includegraphics[width=0.82\linewidth]{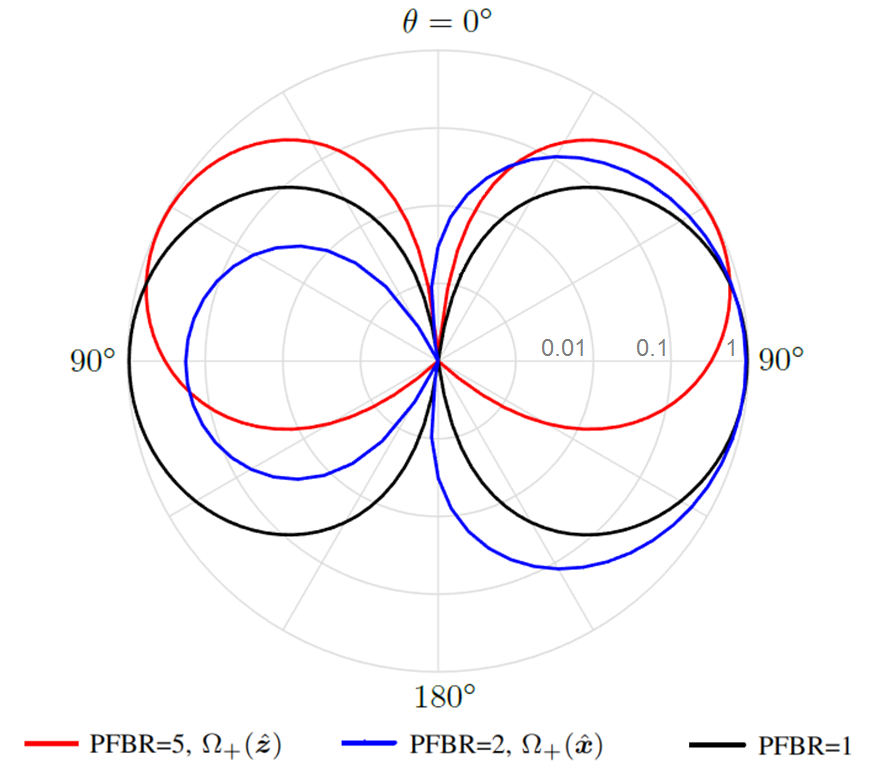}
   \caption{The 2D radiation patterns (normalized in logarithmic scale) for different $\Omega_+(\hu)$ and PFBRs, $xz$-plane.}
  \label{fbr2D}
\end{figure}

For each $\Omega_\pm(\hu)$ and PFBR, we find the associated maximal bandwidth-bound and its associated current distribution and radiation pattern. The radiation patterns for PFBR=5, $\Omega_+(\hz)$ and PFBR=2, $\Omega_+(\hx)$ are shown in Fig. \ref{fbr2D}. We see again that the optimization problem works well on shaping the far-field radiation, for the optimal current. We notice in particular that the radiation pattern with $\Omega_\pm(\hz)$ have interesting possible applications. For instance a downward tilted pattern for PFBR=5, $\Omega_+(\hz)$ in Fig. \ref{fbrQ} is useful e.g. for IoT antennas and WiFi-antennas, since it could reduce the amount of power that is radiated to the sky. 

%!The red curve is rather assymetric, but it depends on $\Omega_+$!  We notice that the red curve in Fig. \ref{fbr2D} is weakly asymmetric about z-axis. That is to say, the asymmetric radiation gives a lower Q than a symmetric. We mention that the associated current with this radiation pattern is an optimal solution of our problem, and there is another optimal current which radiates the mirror image pattern with respect to the z-axis.

The rapid raise of the Q-factor (decrease in bandwidth) for the dipole is associated with the narrow current region $l_x\ll \lambda$ in the previous example. It is thus interesting to investigate a larger current region.

\begin{figure}[htbp]
  \centering
  \includegraphics[width=\linewidth]{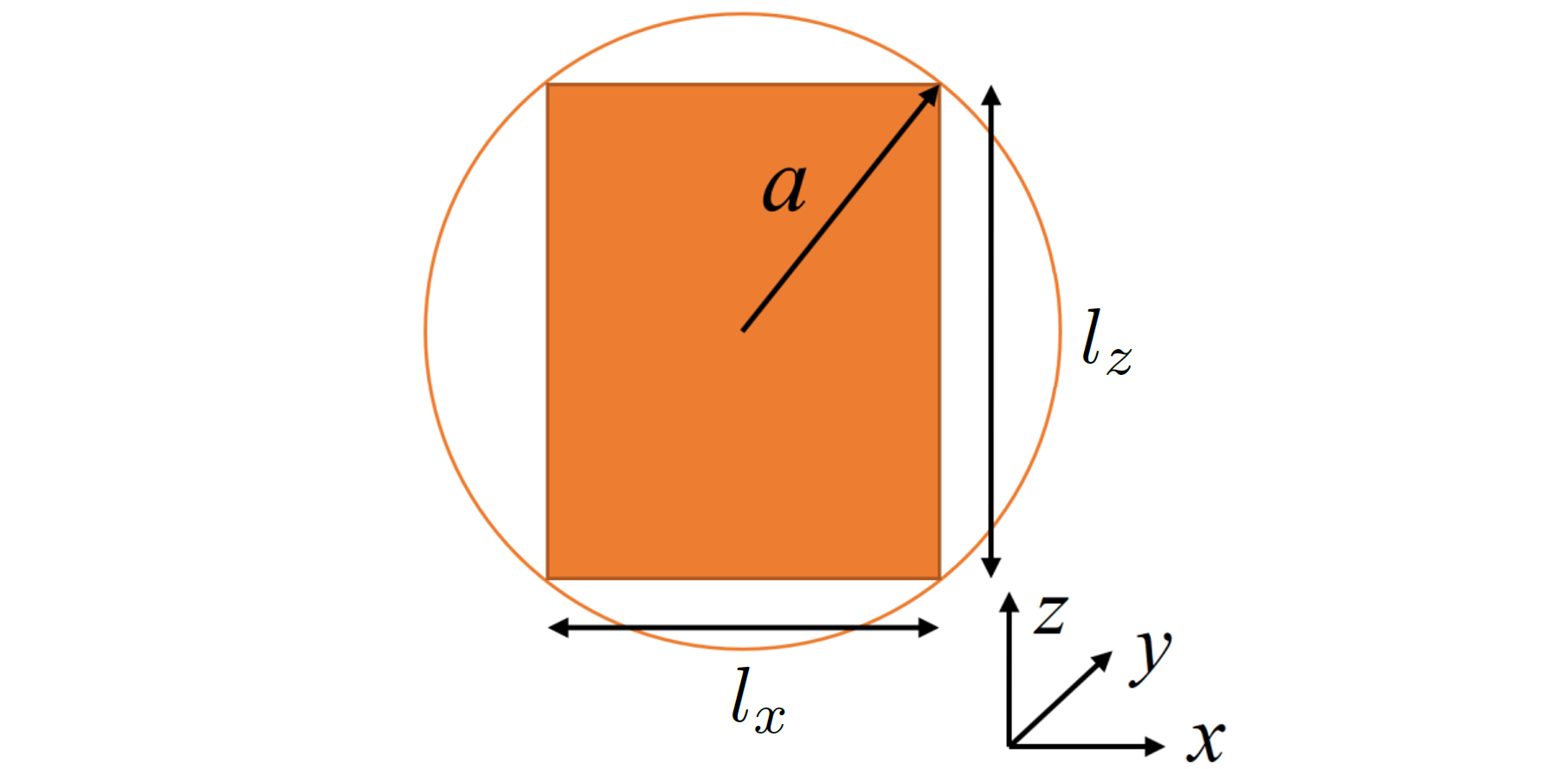}
   \caption{The geometry of the rectangular plates $ka=1$.}
  \label{platesize}
\end{figure}

To do this we study a range of rectangular PEC-plates that fit within $ka=1$. Here $a$ is the radius of a sphere that circumscribes the antenna (also called Chu sphere), see the geometry in Fig. \ref{platesize}. We place the rectangle in the $xz$-plane. We denote the height with $l_z$ and the width $l_x$, for each of the rectangles. 

\begin{figure}[htbp]
  \centering
  \includegraphics[width=\linewidth]{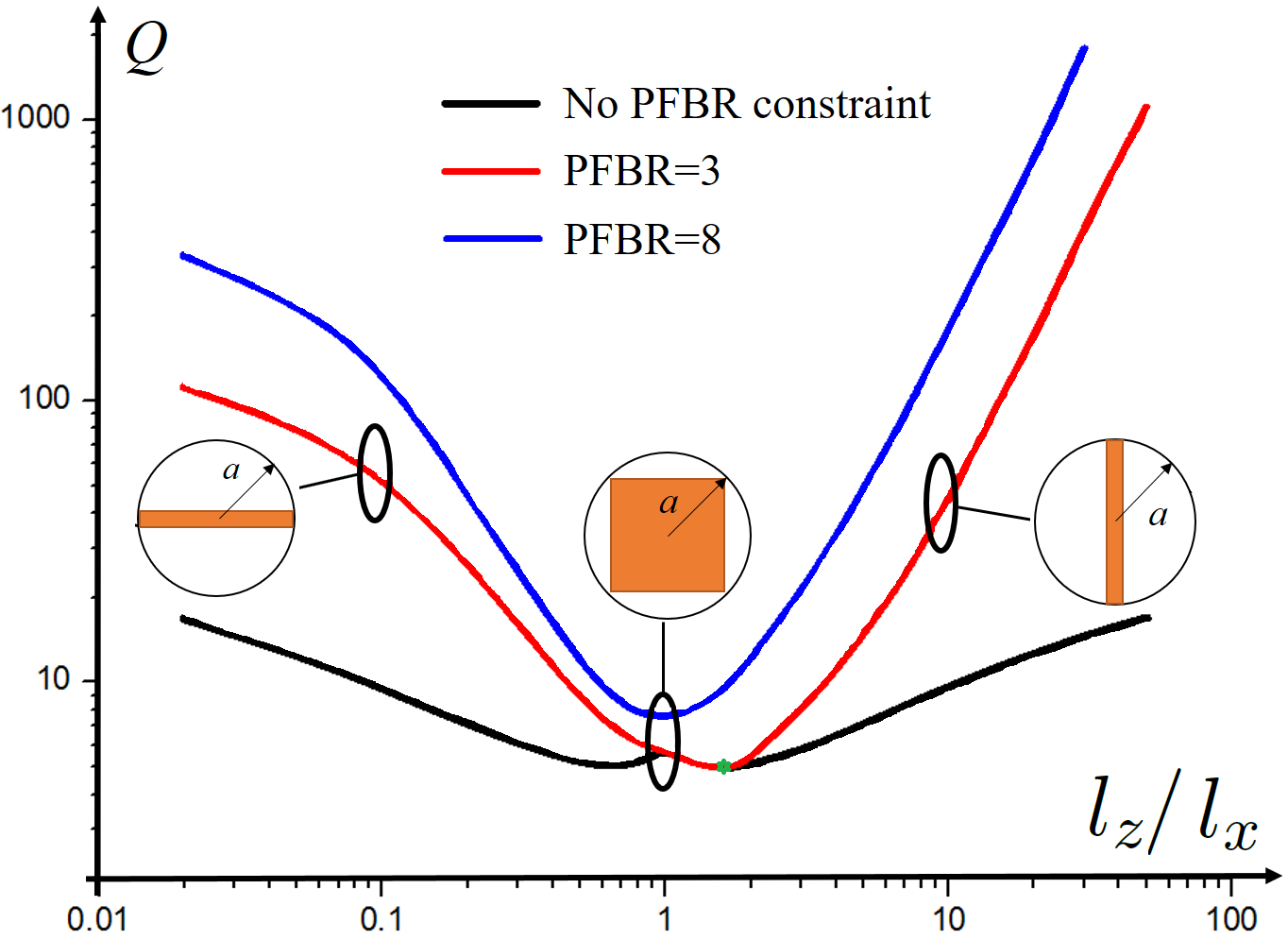}
   \caption{The lower bounds on the Q-factor for rectangular plates $ka=1$, when $\Omega_+(\hx)$ PFBR=3, PFBR=8, and without PFBR constraint.}
  \label{platesizeQ}
\end{figure}

\begin{figure}[htbp]
  \centering
  \includegraphics[width=\linewidth]{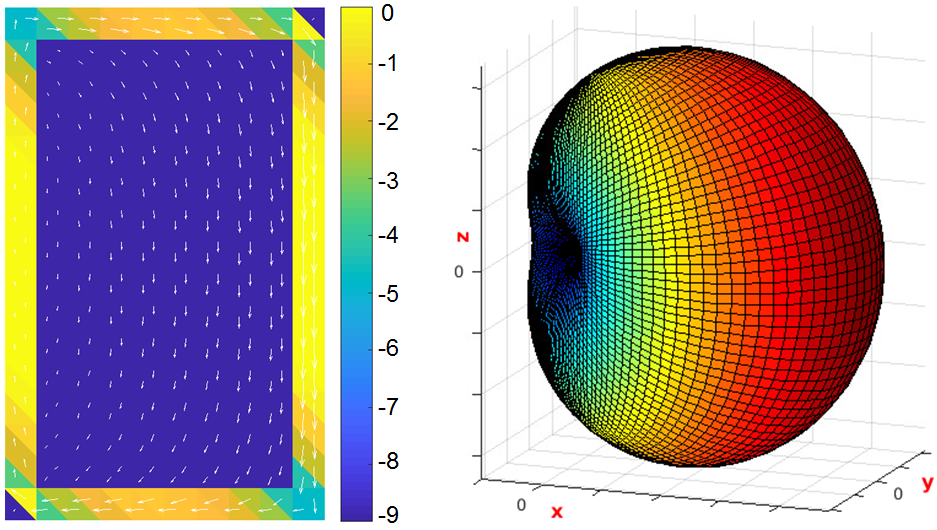}
   \caption{The optimal current distribution (normalized magnitude in dB) on the plate ${l_z}/{l_x}=1.6$,  for $\Omega_+(\hx)$ PFBR=3, and its corresponding radiation pattern. }
  \label{platesize3Dcurrent}
\end{figure}

In Fig.~\ref{platesizeQ} we determine the lowest Q-factor for three different values of the PFBR=$\{1,3,8\}$ while sweeping the rectangular side-ratio $l_z/l_x$ in the interval $[10^{-2},10^2]$. 
As expected, for the same ${l_z}/{l_x}$, a higher PFBR is associated with a higher Q-factor. For a given PFBR constraint, the Q-bound varies with the geometry as depicted. With $\Omega_+(\hx)$ PFBR constraints, the Q-value goes up more rapidly when ${l_z}/{l_x}$ increasing than decreasing associated with the Q-factor cost to generate a higher PFBR. If there is no PFBR constraint, it becomes a minimizing Q problem. As expected, the black curve is symmetric around $l_z/l_x=1$. We can see that we get the lowest Q-value for $\Omega_+(\hx)$ PFBR=3 when ${l_z}/{l_x}\approx 1.6$ (if we only consider ${l_z}\geq {l_x}$), which agrees with the similar result on $D/Q$ in~\cite{Gustafsson2009}. For ${l_z}/{l_x}=1.6$ and $\Omega_+(\hx)$ PFBR=3, we plot the current and radiation pattern in Fig. \ref{platesize3Dcurrent}. In the current plotting, the color shows the normalized magnitude in dB, and the arrows indicate the current distribution at a certain moment. In this case, we get rank one solution to \eqref{PFBR}. From the $xy$-symmetry plane of the problem there should be a second solution, but it is absent here due to the asymmetry of the mesh.

\section{Directivity for an Array Case}\label{sec:Dir}

Gain and directivity are essential antenna design parameters, and it is hence essential to investigate their relation to bandwidth. That infinite directivity can be obtained theoretically has been shown in~\cite{Oseen1922,Uzkov1946,Riblet1948}, the price for a high directivity is a high Q-factor. Efforts to define a boundary between the normal and the superdirective region have been investigated in~\cite{Harrington1958,Geyi2003,Kildal2017}. While investigations of superdirectivity and their realization are many, see e.g.~\cite{Pigeon+etal2014, Hansen2006, Yaghjian+etal2008,Kim2012}, explicit relation between the best possible bandwidth and the directivity are more rare. Initial investigation~\cite{Lo+Lee1966,Margetis+etal1998} were mainly for arrays. Smaller antennas with requirements for given partial directivity was studied in~\cite{Gustafsson+Nordebo2013,Gustafsson+etal2016a,Jonsson+etal2017b} using convex current optimization. Other approaches to explicitly obtain the trade-off between bandwidth and directivity include the eigenvalue methods~\cite{Capek+etal2017,Jonsson+etal2017b,Fabien+etal2017}, and the degrees of freedom (DoF) method~\cite{Kildal2017}.

The directivity $D(\hr)$ is the ratio of the radiation intensity in a direction $\hr$ to the average radiation intensity~\cite{IEEEstandard}:
\begin{equation}\label{Ddef}
D(\hr) = \frac{U(\hr)}{\bar{U}} = \frac{4\pi U(\hr)}{\rP}.
\end{equation}
The antenna gain is defined as $G(\hr)=\eta_\text{eff}D(\hr)$, where $\eta_\text{eff}=\rP/P_{\text{in}}$ is the antenna efficiency. For lossless antennas, we have $G(\hr)=D(\hr)$. Substituting~\eqref{PradI} and~\eqref{UI} into~\eqref{Ddef}, we have the directivity $D(\hr)$ can be represented as:

\begin{equation}\label{DI}
D(\hr) = \frac{4\pi U(\hr)}{\rP} \approx \frac{4\pi}{\eta}\frac{\mathbf{I^H}\big(\mathbf{F^HF}\big)\mathbf{I}}{\mathbf{I^H}\mathbf{R}\mathbf{I}}.
\end{equation}
To design an antenna with directivity of at least $D_0$ in the $\hr$ direction, \ie $D(\hr)\geq D_0$, we have the optimization problem with two quadric constraints~\cite{Jonsson+etal2017b}:

\begin{equation}\label{D}\begin{aligned}
\minimize_{\mathbf{I}\in \CC^N} & \max\{\mathbf{I^H}\mathbf{X_\mathrm{e}}\mathbf{I},\mathbf{I^H}\mathbf{X_\mathrm{m}}\mathbf{I}\big\}\\
\subjectto & \mathop{\mathbf{I^H}\mathbf{R}\mathbf{I}}=1,\\
& \mathop{\mathbf{I^H}\big(\mathbf{F^HF}\big)\mathbf{I}}\geq\frac{\eta D_0}{4\pi}.
\end{aligned}\end{equation}

\begin{figure}[htbp]
  \centering
  \includegraphics[width=\linewidth]{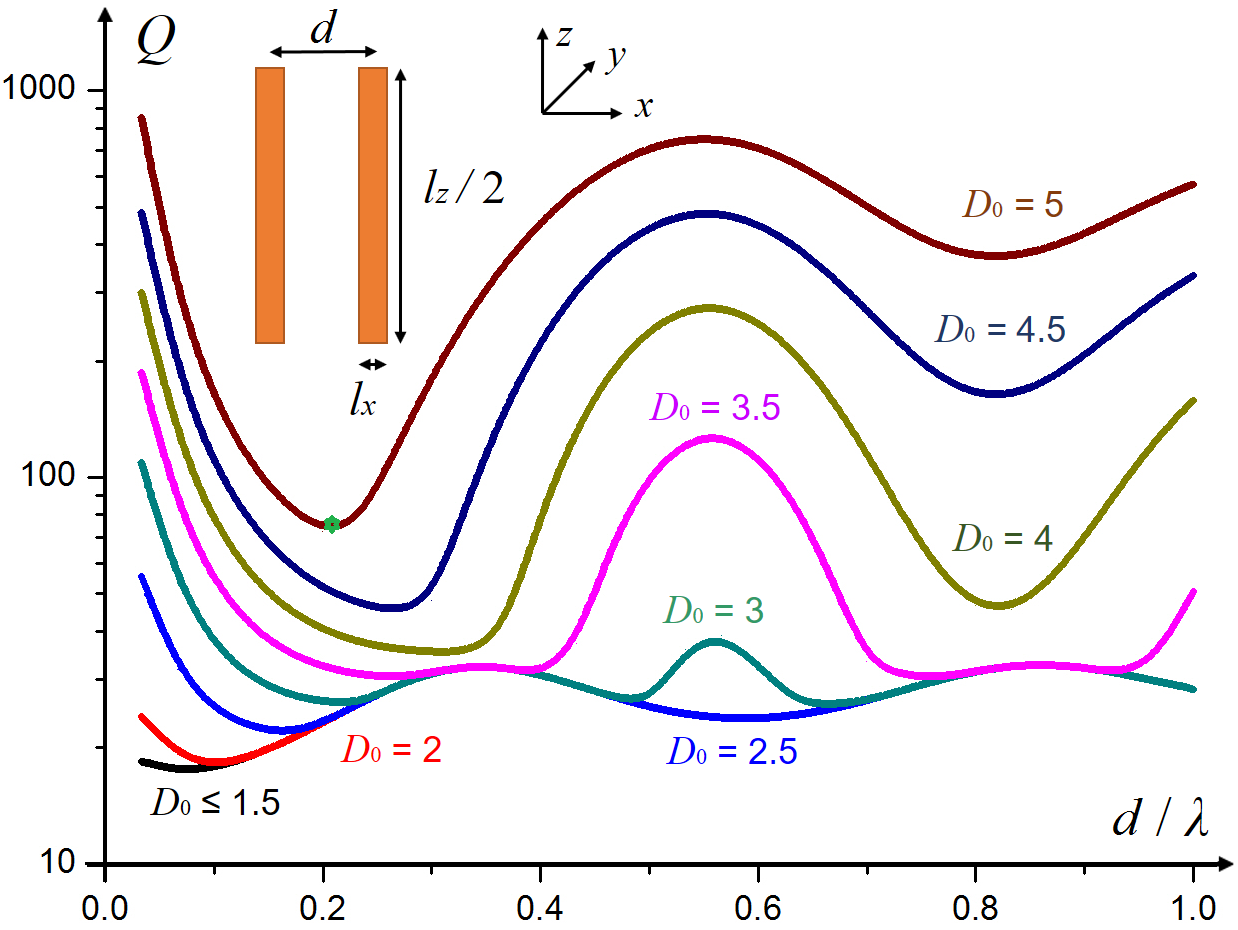}
   \caption{The lower bounds on the Q-factor for different $D(\hx)\geq D_0$ for the two short dipoles structure.}
  \label{2elements}
\end{figure}

The problem \eqref{D} has previously been stated in \cite{Jonsson+etal2017b}, but we return to it here for an interesting array case. Thus let's examine a case with two short dipoles at a distance $d$ from each other along the $x$-direction, see the inset in Fig. \ref{2elements}. In this case it is interesting to compare how the inter-element distance $d$ interact with increased demands on a minimum directivity $D_0$ in the $\hx$ direction. The elements have size of length of $l_z/2=0.24\lambda$ and width of $l_x=0.02l_z$. We find as expected that the lowest possible Q-factor depend on the inter-element distance for a given $D_0$, see the result in Fig. \ref{2elements}.

\begin{figure}[htbp]
  \centering
   \includegraphics[width=\linewidth]{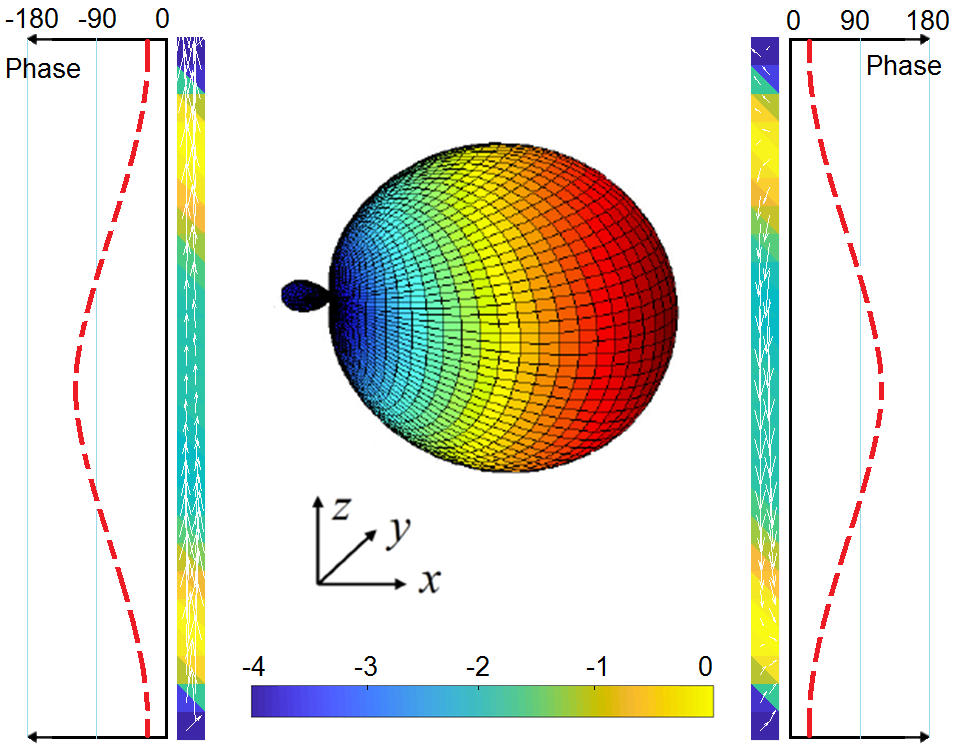}
   \caption{The optimal current distribution on the elements the array, when $d=0.2\lambda$, $D_0=5$, and its corresponding radiation pattern. The current is normalized and the magnitude is expressed in dB and plotted in color on the structure, and the phase of each element is shown in left and right insets respectively.}
  \label{2elements3D}
\end{figure}

For $D_0\leq 1.5$, we have the same bound on Q-factor for $d \in (0, \lambda)$, which means the two dipoles structure has a directivity $D\geq 1.5$, as expected for a small dipole antenna. We observe that the optimal Q-factor depends strongly on both $D_0$ and $d$. Recalling the design rules of Yagi-Uda antennas see e.g.~\cite{Stutzman2012}, we note the expected behavior that Q decrease for certain distances for a fixed $D_0$. For $D_0=5$ we find the first $d$ with a local minimum Q at $d\approx0.2\lambda$. We can think of the two small dipoles as a method to excite certain spherical modes. We conclude that the minimum is a locally better position to excite a high directivity combination of vector-spherical modes. We observe that for $D_0=4$ that a local Q-minimum occurs at $d\approx 0.27\lambda$ while for $D_0=5$ we have minimum at $d\approx 0.2\lambda$. The normalized current distribution is depicted in Fig. \ref{2elements3D} in dB scale and the radiation pattern for $D_0=5$, $d=0.2\lambda$ are also shown.

\section{Antenna Design}\label{sec:Design}

In the above sections we have concentrated on finding the optimal Q-factor under different set of constraints. The above theory is naturally independent of the antenna shape, but we have here deliberately concentrated on the dipole structure, to simplify efforts towards the realization of the desired patterns. In this section we concentrate in particular on two of the above cases: the flattening of the radiation pattern, `beam-shaping', as investigated in Section \ref{sec.Beam}, see Fig. \ref{flat3D}b, and the upward-tilting of the radiation pattern as a result of a desired PFBR for $\Omega_\pm(\hz)$, Section \ref{sec.PFBR}, see Fig. \ref{fbrQ}. In this section we aim to realize these radiation patterns with low Q-factor. The design idea is to use a multi-position feeding strategy for the element. Similar ideas have been studied for other cases in see e.g.~\cite{Jelinek+Capek2017}.  

\begin{figure}[htbp]
  \centering
  \includegraphics[width=\linewidth]{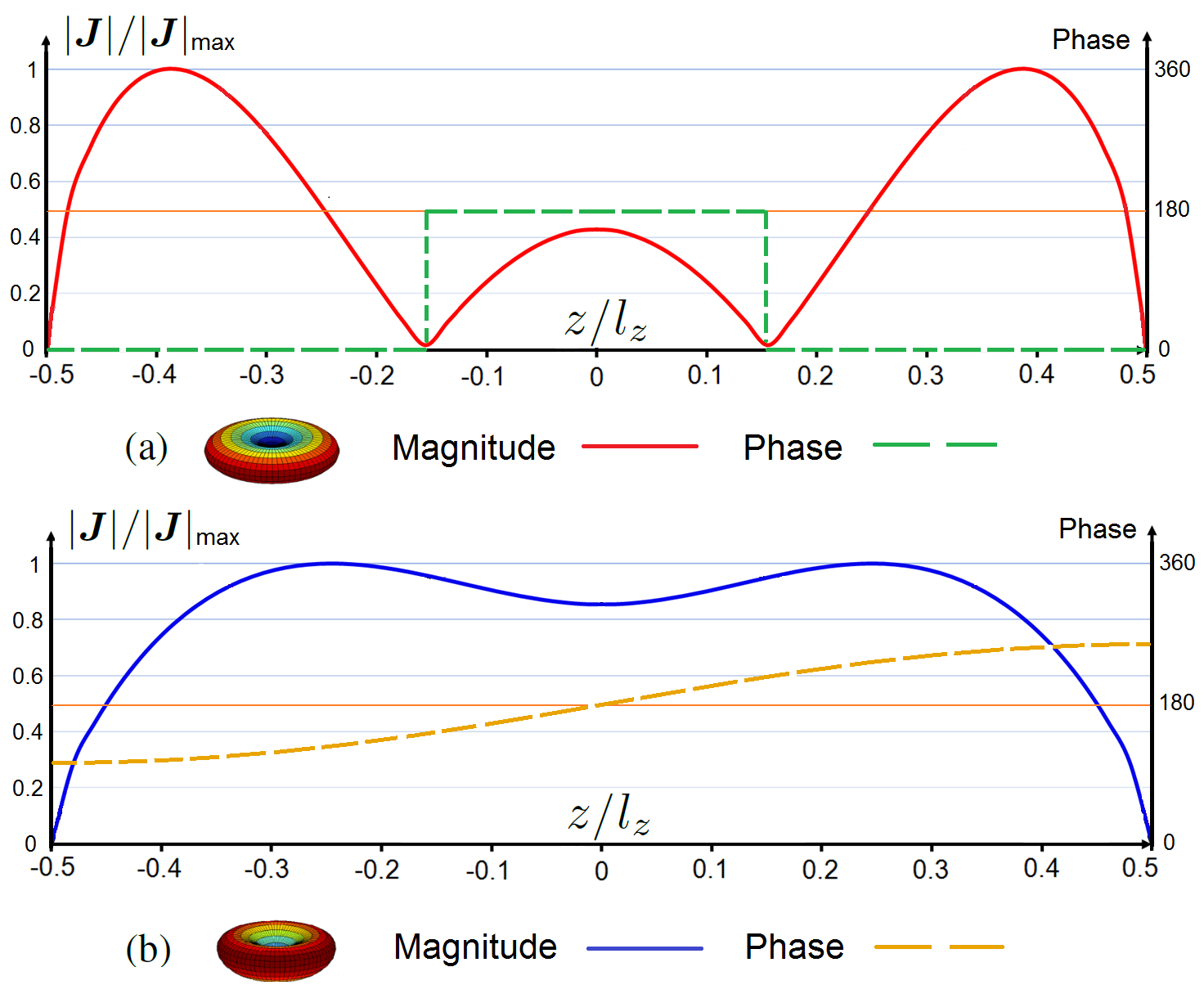}
   \caption{The optimal current distributions (with normalized magnitude and phase) for the dipole with different radiation patterns. (a) is the optimal current for a beam-shaping pattern, with a narrower pattern, the same case as given in Fig. \ref{flat3D}b. (b) is a PRBR-case with a tilted pattern corresponding to the $\Omega_+(\hz)$-case in Fig. \ref{fbrQ}. }
  \label{dipolecurrent}
\end{figure}

In the above optimization procedures we have investigated different meshing for the considered structures, and the associated current does to some small degree depend on the choice of mesh. In the two selected cases we can approximate the optimal current with a one-dimensional current as depicted in Fig. \ref{dipolecurrent}. Here the surface current density is normalized with the maximum amplitude of the current. 

We use a three port feeding strategy to strive to realize two antennas with the respective desired radiation pattern: a narrow sector beam pattern, (a), and a tilted pattern, (b), in Fig.~\ref{dipolecurrent}. In the usual MoM-way, the gap feeding voltages are related to the current on the structure through the impedance matrix $\mathbf{Z}$ as $(\mathbf{ZI})|_\text{port} = \mathbf{V}_\text{port}$. Thus we can optimize the complex feeding by rewriting the minimizing Q problems \eqref{Shaping} and \eqref{PFBR} with only exciting voltage (or current) at the ports in three positions, denoted $\Sigma_1$, i.e., $\mathbf{V_\text{port}}= \mathbf{v}$, and for all other RWG-edges, $\Sigma_2=S\backslash \Sigma_1$, we set the elements of $\mathbf{V}$ to zero. This implies that we can decompose the current vector into two parts $\mathbf{I^T}=[\mathbf{I_\mathrm{1}^T \; I_\mathrm{2}^T}]$, associated with the respective regions $\Sigma_1,\Sigma_2$. %\SHs{correspond to the region with $V=0$ of the antenna.} 
%\LJ{((Port current is a bit unclear to me, we have port/gap voltages, it does not mean that we trivially have the "port current" since $I_1$ and $I_2$ couple through the reduced Z-matrix))} 
According to the EFIE, we have the relation
\begin{equation}
\mathbf{Z_\mathrm{21}}\mathbf{I_\mathrm{1}}+\mathbf{Z_\mathrm{22}}\mathbf{I_\mathrm{2}}=\mathbf{0},
\end{equation}
and then we can express the induced current $\mathbf{I}_\mathrm{2}$ with $\mathbf{I}_\mathrm{1}$:
\begin{equation}\label{induced}
\mathbf{I_\mathrm{2}}=-\mathbf{Z_\mathrm{22}}^{-1}\mathbf{Z_\mathrm{21}}\mathbf{I_\mathrm{1}} =\tilde{\mathbf{Z}}\mathbf{I_\mathrm{1}}.
\end{equation}
The minimizing Q problems \eqref{Shaping} and \eqref{PFBR} can be rewritten into a quadratic form over a small matrix by replacing $\mathbf{I}$ with $\mathbf{I_\mathrm{1}}$, and by replacing $\mathbf{X_e}$, $\mathbf{X_m}$ and $\mathbf{R}$ with $\mathbf{\tilde{X}_e}$, $\mathbf{\tilde{X}_m}$ and $\mathbf{\tilde{R}}$, where

\begin{equation}
\mathbf{\tilde{X}}=\mathbf{X_\mathrm{11}}+\mathbf{X_\mathrm{12}}\mathbf{\tilde{Z}}+\mathbf{\tilde{Z}^H}\mathbf{X_\mathrm{21}}+\mathbf{\tilde{Z}^H}\mathbf{X_\mathrm{22}}\mathbf{\tilde{Z}},
\end{equation}
\begin{equation}
\mathbf{\tilde{R}}=\mathbf{R_\mathrm{11}}+\mathbf{R_\mathrm{12}}\mathbf{\tilde{Z}}+\mathbf{\tilde{Z}^H}\mathbf{R_\mathrm{21}}+\mathbf{\tilde{Z}^H}\mathbf{R_\mathrm{22}}\mathbf{\tilde{Z}}.
\end{equation}
This approach is similar to the current optimization problems for embedded antennas and arrays, and a longer discussion can be found in e.g.~\cite{Gustafsson+Nordebo2013,Jonsson2017,Jonsson2018}.

\begin{figure}[htbp]
  \centering
  \includegraphics[width=\linewidth]{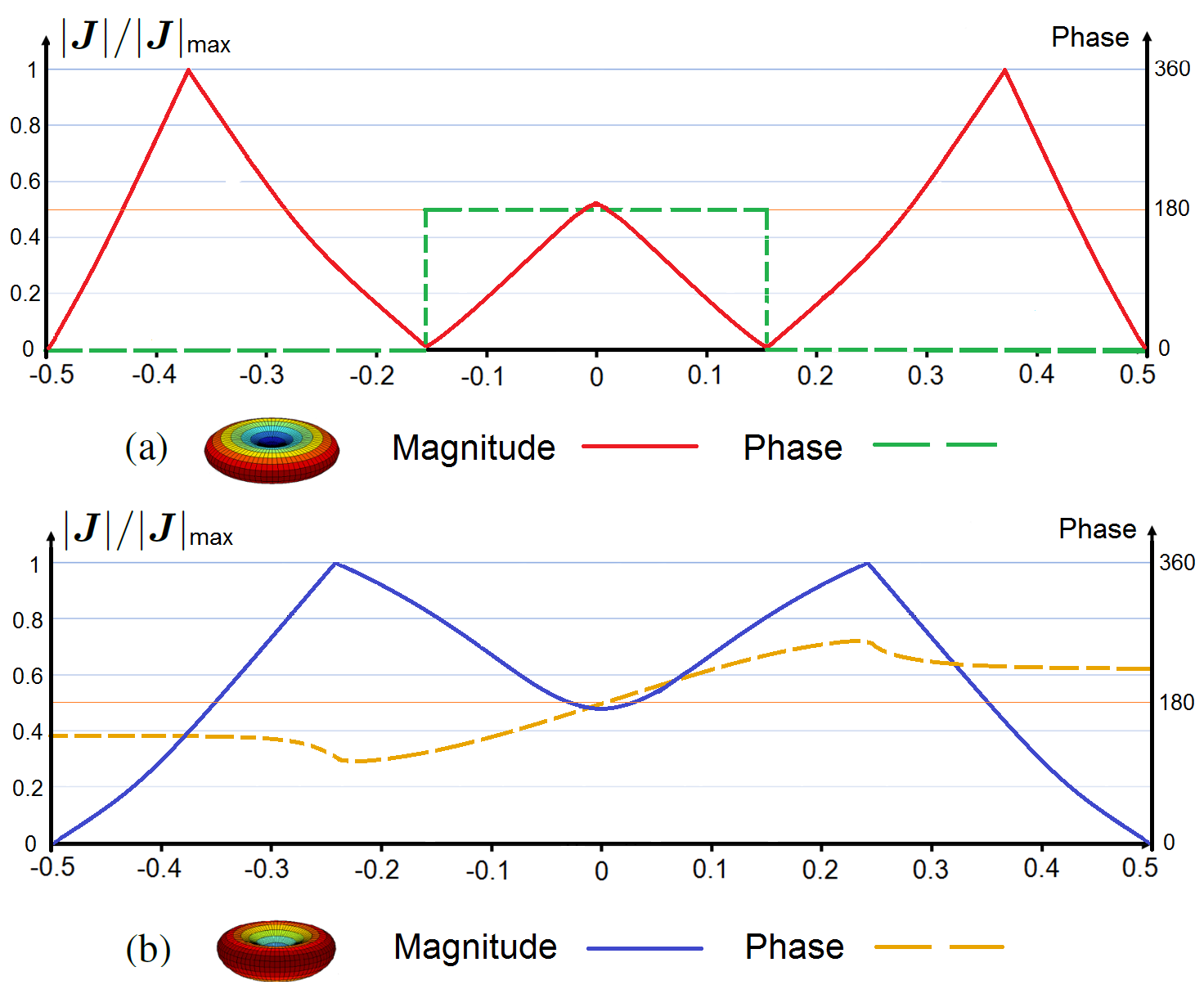}
   \caption{Current distributions (with normalized magnitude and phase) for the dipole by optimal multi-position feedings, with different radiation patterns. (a) is the current for a beam-shaping pattern, with a narrower pattern, the same with Fig. \ref{flat3D}b. (b) is a PRBR-case with a tilted pattern corresponding to the $\Omega_+(\hz)$-case in Fig. \ref{fbr2D}.}
  \label{dipolecurrent2}
\end{figure}

The here used strategy is to use the current distributions in Fig. \ref{dipolecurrent} to select the feeding positions at its crests and troughs: $z=[-0.39l_z, 0, 0.39l_z]$ for the narrow sector beam (b), and $z=[-0.27l_z, 0, 0.27l_z]$ for the tilted pattern (c). The reduced optimization problem determines the optimal excitation of the dipole antennas at each port, satisfying the far-field constraint and determines the corresponding Q-factor. Once the excitation is known, we use~\eqref{induced} to determine the surface currents for the multi-position feeding case, which is inserted into~\eqref{Qdef} to obtain the Q-factor. The resulting surface current distribution is depicted in Fig. \ref{dipolecurrent2}. Please observe that the three-port feeding structure give surface currents that are roughly similar in shape to the ideal currents shown in Fig.~\ref{dipolecurrent}. 

\begin{figure}[htbp]
  \centering
  \includegraphics[width=\linewidth]{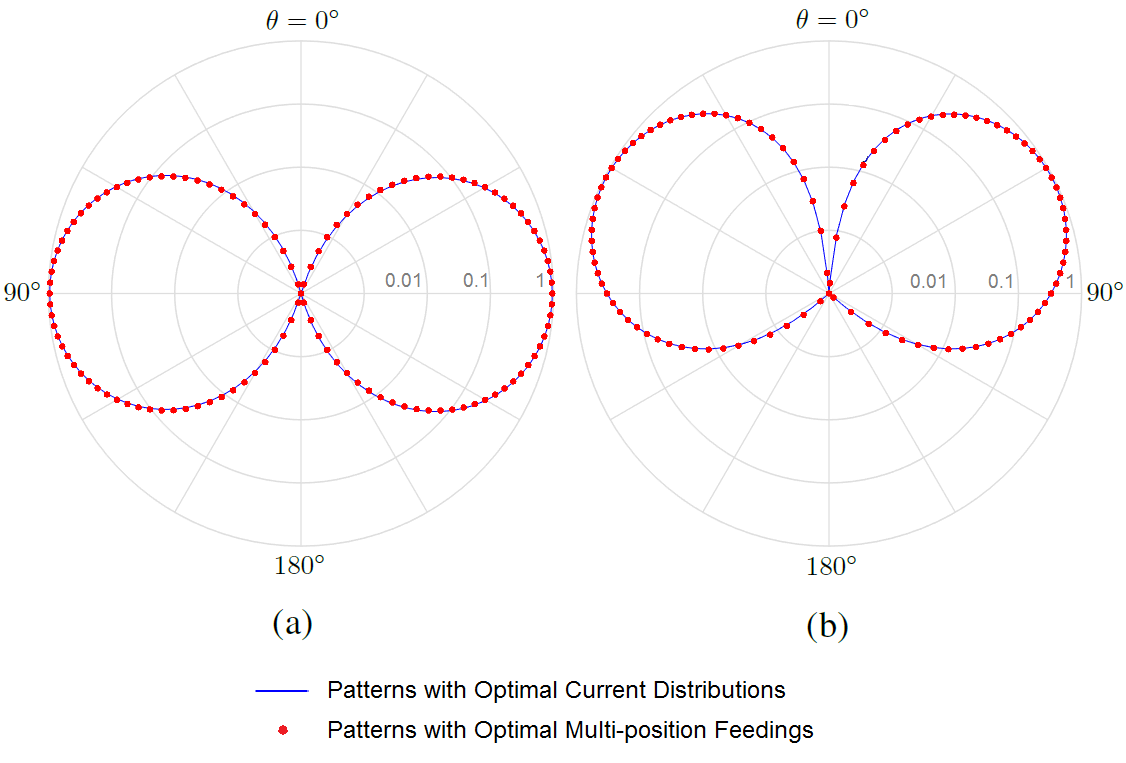}
   \caption{The 2D radiation patterns (normalized in logarithmic scale) comparison between by optimal current distributions and by optimal multi-position feedings. (a) is a narrower dipole pattern , the same with Fig. \ref{flat3D} (b). (b) is a tilted pattern, the same with the $\Omega_+(\hz)$ case in Fig. \ref{fbr2D}.}
  \label{2Dfeed}
\end{figure}

The associated radiation pattern for the multi-position feedings and the original cases are depicted in Fig. \ref{2Dfeed}. We observe that the agreement between the current-optimized radiation pattern and the multi-position feedings optimized radiation patterns are very similar, the deviation is almost not visible. If we turn to the Q-factor, we notice that the multi-position feeding version of the narrow-radiation pattern (beam-forming) became Q=201, as compared with 125, i.e., the multi-position feeding strategy to beam-forming realize the desired pattern, but the bandwidth has reduced. For the second case, the multi-position feeding strategy for the beam-tilting-case have a Q-factor of 46 whereas the current optimization has Q=33. Both Q-factors/bandwidth are included for comparison in the parameters sweep, see the pink point in Fig. \ref{flatQ} and the green point in Fig. \ref{fbr2D}. For both cases of multi-position feeding we have that $Q_\text{antenna}\leq 1.61Q_\text{optimal}$.

Above we have illustrated that it is possible with rather simple methods to realize the desired radiation patterns, but at the cost of a higher Q-factor, and the complication of a three-position feeding strategy. For the beam-shaping associated with the $\Omega_+(\hx)$-case we note that the Q-factor is very high even for a small PFBR, and we hence expect that it is harder to realize this case. For the two element Directivity case a six port feeding strategy was tried, however in this effort the Q-factor turned out to be too high as compared with the optimum for us to include the case here, and a different strategy is required to realize antennas close to the bound. Such a strategy for a larger array case is considered in~\cite{Jonsson2018}. 

\section{Conclusion}

In this paper, we illustrate how the SDR optimization technique opens up for several new types of far-field constraints. We introduce and test two new far-field constraints, i.e., power beamwidth and PFBR, and we also revisit the super-directive constraints this time for a small array. We have investigated how a multi-element antenna can influent directivity as a function of the inter-distance between the elements, and provide a tool to predict the Q-factor as a function of distance and desired directivity.  

The optimization problems here are all limited to a maximum of two additional constraints, and our results are of rank one, thus the relaxation method yields a solution to the original non-relaxed problem. It is interesting to observe that also rather small and narrow antennas can have a non-standard radiation pattern, with a small extra cost in Q-factor. 

The multi-position feeding strategy investigated here increases our understanding of realization of optimal antennas. In particular, we observe that we have essentially the optimal predicted far-fields, but that the Q-factor deviates, this is due to the smoothing effect that the radiated field as a function of the current, and as a contrast to the sensitivity of the reactive field. This strategy then translates into an expected cost in bandwidth (Q-factor). It also illustrates that these non-standard radiation patterns discussed in the first part of this paper can be realized in certain cases, while the Q-factor remains reasonably low.

\section*{Acknowledgements}

We gratefully acknowledge the support of the Swedish Foundation for Strategic Research for the project "Convex analysis and convex optimization for EM design" (SSF/AM130011), and the Swedish Governmental Agency for Innovation Systems through the Center ChaseOn in the project "iAA" (ChaseOn/iAA). Mr. S. Shi is supported by the China Scholarship Council (CSC), which is gratefully acknowledged.

%\bibliographystyle{IEEEtran}
%\bibliography{total,add}
%\input{TrDir_w.bbl}


\bibliographystyle{IEEEtran}
\bibliography{IEEEabrv,shuaibibfile}

\end{document}